\documentclass[11pt,a4paper,fleqn]{article}

\usepackage{amsmath,amssymb,amsthm,enumerate,cite}
\usepackage{longtable}

\newcommand{\ev}{\mathop{\mathbf{ev}}\nolimits}
\newcommand{\F}{\mathbb{F}}
\newcommand{\LnF}{\mathop{\mathcal L_n(\mathbb F)}}

\newcommand{\GlnF}{\mathop{{\rm GL}(n,\F)}\nolimits}
\newcommand{\GlF}{\mathop{{\rm GL}(3,\F)}\nolimits}

\newcounter{mcasenum}

\newtheorem{theorem}{Theorem}[section]

\newtheorem*{lemma*}{Lemma}

\newtheorem*{corollary*}{Corollary}

{\theoremstyle{definition} 
\newtheorem*{definition*}{Definition}

\newtheorem*{example*}{Example}
\newtheorem*{examples*}{Exampes}
\newtheorem{remark}[theorem]{Remark}
\newtheorem*{remark*}{Remark}

\newtheorem*{result*}{Result}
\newtheorem*{comment*}{Comment}

\begin{document}

\title{\bf Describing certain Lie algebra orbits\\ via polynomial equations
}

\author{N.M. Ivanova$^a$ and C.A. Pallikaros$^{b}$}

\date{July 28, 2017}

\maketitle

{\vspace{1mm}\par\noindent\footnotesize\it{
${}^{a}$~Institute of Mathematics of NAS of Ukraine,~3 Tereshchenkivska Str., 01601 Kyiv, Ukraine\\
\phantom{${}^{a}$~}{}E-mail: ivanova.nataliya@gmail.com\\
${}^{b}$~Department of Mathematics and Statistics, University of Cyprus, PO Box 20537,
1678 Nicosia, Cyprus\\
\phantom{${}^{b}$~}{}E-mail: pallikar@ucy.ac.cy\\

}}

\begin{abstract}
Let $\mathfrak{h}_3$ be the Heisenberg algebra and let $\mathfrak g$ be the 3-dimensional Lie algebra having $[e_1,e_2]=e_1\,(=-[e_2,e_1])$ as its only non-zero commutation relations. We describe the closure of the orbit of a vector of structure constants corresponding to $\mathfrak{h}_3$ and $\mathfrak g$ respectively as an algebraic set giving in each case a set of polynomials for which the orbit closure is the set of common zeros.
Working over an arbitrary infinite field, this description enables us to give an alternative way, using the definition of an irreducible algebraic set, of obtaining all degenerations of $\mathfrak{h}_3$ and $\mathfrak g$ (the degeneration from $\mathfrak g$ to $\mathfrak{h}_3$ being one of them).
\\[1.5ex]
\end{abstract}

\section{Introduction}

In the second half of the twentieth century a lot of works appeared on the study of different types of limit processes between
various physical or geometrical theories.
Such limit processes naturally lead to the notion of contraction (or degeneration).
Possibly the first work in this direction was Segal~\cite{Segal1951}
who considered the non-isomorphic limit of sequences of structure constants of some isomorphic Lie groups.
Such limit processes are called contractions.
The claim is that if two physical theories are related by a limit process,
then the associated invariance groups (and invariance algebras) should also be related by some limit process.
This led to a wide investigation of contractions of Lie algebras from the physical point of view.
Possibly, the three most famous physical examples of contractions are the following.

\begin{itemize}

\item
Contraction of relativistic mechanics to classical mechanics was studied in works by
In\"on\"u and Wigner~\cite{InonuWigner1953,InonuWigner1954}.
Considering the physical limit process $c\to\infty$ in special relativity theory
they showed how the symmetry group of relativistic mechanics
(the Poincar\'e group) contracts to the Galilean group which is the symmetry group of classical mechanics.

\item The relation between classical and quantum mechanics can also be expressed in terms of a limit process or, in other words, a contraction~\cite{Hepp1974}.
Thus, one can consider classical mechanics as the limit of quantum mechanics
under the contraction $\mathfrak{h}\to\mathfrak{a}$,
where $\mathfrak{h}$ is the Weyl--Heisenberg algebra and $\mathfrak{a}$
is the abelian Lie algebra of the same dimension.
Under this contraction the quantum mechanical commutator $[x,p]=i\hslash$ (corresponding to the Heisenberg uncertainty principle)
maps to the Abelian case (that is, the classical mechanics limit) under $\hslash\to 0$.

\item
The porous medium equation $u_t=m^{-1}\Delta(u^m-1)$ can be contracted~\cite{Wu1993} (as $m\to0$) to the
equation $u_t=\Delta\ln u$, which is equivalent to the equation defining the Ricci flow on~$\mathbb R^2$.
\end{itemize}

In these (and many other publications) it is shown, in particular, how some basic properties of the ``contracted theories'' can be reconstructed from the corresponding properties of the ``original'' theories.
Such observations were summarized by Zaitsev~\cite{Zaitsev1961} who, independently of In\"on\"u and Wigner, suggested constructing ``the theory of  physical theories'' based on group limits of physical theories. This amounts to including in a uniform system several physical theories being connected together via certain relations.

Recently, different types of contractions have been widely used in elementary particle theory, analysis of differential equations and other
areas of mathematical and theoretical physics.

Working over $\mathbb C$ or $\mathbb R$, the statement ``Lie algebra $\mathfrak h_1$ is a contraction of Lie algebra $\mathfrak h_2$'' can be rephrased as ``$\mathfrak h_1$ lies in the closure, in the metric topology, of the orbit of $\mathfrak h_2$ under the `change of basis' action of the group of invertible linear transformations''.
In~\cite{GrunewaldOHalloran1988a} the authors show that over $\mathbb C$ the orbit closure in the metric topology coincides with the orbit closure in the Zariski topology.
Orbit closures with respect to the Zariski topology are called degenerations.
The notion of degeneration is well-defined not only over the fields $\mathbb C$ and $\mathbb R$ but also over an arbitrary ground field.
In fact, this concept of orbit closure  under the action of various groups arises naturally in many areas of mathematics (see, for example,~\cite{Kraft1982}).

In~\cite{IvanovaPallikaros2017} we explored the possibility of investigating degenerations over an arbitrary field using elementary algebraic techniques.
For this we needed to extend or modify techniques already used over the fields $\mathbb C$, $\mathbb R$
(for example contractions obtained as limit points resulting from the action of diagonal matrices, also known as generalized In\"on\"u--Wigner contractions)
in a way so that they can be applied to the case of degenerations over an arbitrary field.
In this paper, although we continue our study of degenerations via an elementary algebraic approach, we take a slightly different path and consider the possibility of obtaining all degenerations (for certain examples of Lie algebras) `from first principles' by direct application of the definition of an algebraic (Zariski-closed) set.
This involves obtaining explicit descriptions of the orbit closures under consideration using polynomial equations.

The paper is organized as follows.
In Section~\ref{SectionGenPrel} we give some necessary background, the setup being over an arbitrary infinite field $\F$.
In particular, in Subsection~\ref{SubsecAlgSets} we recall some basic definitions and results on irreducible algebraic sets and regular maps while in Subsection~\ref{SubsecDegenLieAlgs} we recall the definition of degeneration together with some basic facts on Lie algebra structure vectors and their orbits under the `change of basis' action of the general linear group.
In Section~\ref{SectionDegenExamples} we perform some explicit computations concerning the orbits (and their closure in the Zariski topology) of certain given Lie algebra structure vectors corresponding to $\mathfrak h_3$ and $\mathfrak g_2\oplus\mathfrak a_1$ respectively, where $\mathfrak h_3$  denotes the Heisenberg algebra, $\mathfrak g_2$ denotes the 2-dimensional non-Abelian Lie algebra and $\mathfrak a_1$ denotes the 1-dimensional Abelian Lie algebra.
This enables us to give a description of the orbit closures of these structure vectors as algebraic sets via polynomial equations and, as a consequence, determine in an alternative way all degenerations of $\mathfrak h_3$ and $\mathfrak g_2\oplus\mathfrak a_1$  over $\F$.
We also obtain descriptions of the particular orbits described above as the intersection of a Zariski-closed with a Zariski-open set.

\section{Preliminaries and generalities}\label{SectionGenPrel}

We begin this section by recalling some basic facts on irreducible algebraic sets.
We refer the reader to Geck~\cite{Geck2003} for more details and for proofs of the main results from the theory we will be using.

\subsection{Algebraic sets}\label{SubsecAlgSets}

Fix $\F$ to be an arbitrary infinite field and let $m$ be a positive integer.
We consider the ring $F[\boldsymbol X]=\F[X_1,\ldots,X_m]$ of polynomials in the indeterminates $X_1,\ldots,X_m$ over $\F$.
For each $\boldsymbol\alpha=(\alpha_1,\ldots,\alpha_m)\in\F^m$ there exists a unique $\F$-algebra homomorphism $\ev_{\boldsymbol\alpha}:\F[X_1,\ldots, X_m]\to \F$
such that $\ev_{\boldsymbol\alpha}(X_i)=\alpha_i$ for all $i$.
Given $\boldsymbol\alpha=(\alpha_1,\ldots,\alpha_m)\in\F^m$ and $f\in\F[X_1,\ldots,X_m]$ we will be writing more simply $f(\boldsymbol\alpha)=f(\alpha_1,\ldots,\alpha_m)=\ev_{\boldsymbol\alpha}(f)$.

\begin{definition*}
Let $S$ be any subset of $\F[X_1,\ldots,X_m]$.
The algebraic set ${\bf V}(S)$ determined by $S$ is defined by
\[
{\bf V}(S)=\{\boldsymbol\alpha\in \F^m:\ f(\boldsymbol\alpha)=0\mbox{ for all } f\in S\}.
\]
A subset of $\F^m$ is called \emph{algebraic} if it is of the form ${\bf V}(S)$ for some subset $S\subseteq \F[X_1,\ldots,X_m]$.
For any subset $V\subseteq \F^m$, the vanishing ideal ${\bf I}(V)$ of $V$ is defined by
\[
{\bf I}(V)=\{f\in \F[X_1,\ldots, X_m]:\ f(\boldsymbol\alpha)=0 \mbox{ for all } \boldsymbol\alpha\in V\}.
\]
\end{definition*}

It is immediate from the above definition that if $S_1$, $S_2$ are subsets of $\F[X_1,\ldots,X_m]$ with $S_1\subseteq S_2$, then ${\bf V}(S_2)\subseteq{\bf V}(S_1)$ (see~\cite[Remark 1.1.4]{Geck2003}).

It can be shown (see, for example,~\cite[Remark 1.1.4 and Lemma 1.1.5]{Geck2003} that arbitrary intersections and finite unions of algebraic sets in $\F^m$ are again algebraic.
The empty set $\varnothing$ and $\F^m$ itself are clearly algebraic.
Thus, the algebraic sets in $\F^m$ form the closed sets of a topology in $\F^m$, which is called the \emph{Zariski topology}.
A subset $X\subseteq \F^m$ is open if its complement $\F^m\setminus X$ is algebraic (closed).

We will denote by $\overline{V}$ the closure of a subset $V$ of $\F^m$ in the Zariski topology.

\bigskip

An essential role in our investigation is played by the notion of irreducibility of algebraic sets.

\begin{definition*}
Let $Z\subseteq\F^m$ be a nonempty algebraic set.
We say that $Z$ is \emph{reducible} if we can write $Z=Z_1\cup Z_2$, where $Z_1,Z_2\subseteq Z$ are nonempty algebraic subsets with $Z_1\ne Z$ and $Z_2\ne Z$.
Otherwise, we say that $Z$ is \emph{irreducible}.
\end{definition*}

\begin{remark}[{see \cite[Example 1.1.13]{Geck2003}}]\label{Remark1.1.13Geck}
Our assumption that $\mathbb{F}$ is infinite ensures that $\mathbb{F}^m$ is irreducible.
\end{remark}

\begin{definition*}\label{DefinitionRegulMap}
Let $s,r$ be positive integers and let $V\subseteq \mathbb{F}^s$ and $W\subseteq \mathbb{F}^r$ be nonempty algebraic sets.
We say that $\Phi:V\to W$ is a regular map if there exist $f_1,\ldots,f_r\in\mathbb{F}[X_1,\ldots,X_s]$ such that $\Phi(\boldsymbol\alpha)=(f_1(\boldsymbol\alpha),\ldots,f_r(\boldsymbol\alpha))$ for all $\boldsymbol\alpha\in V$.
\end{definition*}

One can then observe (see~\cite[page 23]{Geck2003}) that regular maps are continuous in the Zariski topology.

\begin{remark}[{see \cite[Remark 1.3.2]{Geck2003}}]\label{Remark1.3.2Geck}
Let $V$, $W$ be as in the definition above and let $\Phi:V\to W$ be a regular map.
Assume that $V$ is irreducible.
Then the Zariski closure $\overline{\Phi(V)}\subseteq W$ is also irreducible.
\end{remark}

\subsection{Degenerations of Lie algebras}\label{SubsecDegenLieAlgs}

We keep the setup of the previous subsection.
In particular $\F$ denotes an arbitrary infinite field but now we assume further that $m=n^3$ for some integer $n\ge2$ we have fixed.
Also let $G$ be the general linear group $\GlnF$.

Now let $\boldsymbol\alpha=(\alpha_1,\ldots,\alpha_m)\in\F^m$ be given.
For the rest of our discussion, it will be convenient to relabel the components of $\boldsymbol\alpha$ as follows.
For $1\le r\le m$ relabel $\alpha_r$ as $\alpha_{i(r),j(r),k(r)}$ where $i(r)$, $j(r)$, $k(r)$ are the unique integers with $1\le i(r),j(r),k(r)\le n$ satisfying $r-1=(i(r)-1)n^2+(j(r)-1)n+(k(r)-1)$.
We will be writing $\boldsymbol\alpha=(\alpha_{i,j,k})$ or $\boldsymbol\alpha=(\alpha_{ijk})$ for short.
For example, in the case $n=2$ ($m=8$) we have for $\boldsymbol\alpha\in\F^m$,
\[
\boldsymbol\alpha=(\alpha_1,\alpha_2,\alpha_3,\alpha_4,\alpha_5,\alpha_6,\alpha_7,\alpha_8)= (\alpha_{111},\alpha_{112},\alpha_{121},\alpha_{122},\alpha_{211},\alpha_{212},\alpha_{221},\alpha_{222}).
\]
(The above ordering in fact amounts to writing $\boldsymbol\alpha=(\alpha_{ijk})\in\F^{n^3}$ where the triples $(i,j,k)$ are placed in lexicographic order.)

In a similar manner we relabel the indeterminates $X_1,\ldots,X_m$ in $\F[X_1,\ldots,X_m]$ and we write $\F[\boldsymbol X]\,(=\F[X_1,\ldots,X_m])= \F[X_{ijk}:\ 1\le i,j,k\le n]$.

\begin{definition*}
An element $\boldsymbol\lambda=(\lambda_{ijk})\in\F^m$ is called a Lie algebra structure vector if there exists an $n$-dimensional Lie algebra $\mathfrak{g}$ over $\F$ and an ordered  $\F$-basis
$\hat b=(b_1,\ldots,b_n)$ of $\mathfrak{g}$ such that $[b_i,b_j]=\sum_{k=1}^n\lambda_{ijk}b_k$ for $1\le i,j\le n$.
In such a case we call $\boldsymbol\lambda=(\lambda_{ijk})$ the structure vector of $\mathfrak{g}$ relative to $\hat b$.
We denote by $\LnF$ the subset of $\F^m$ consisting of precisely those elements of $\F^m$ which are Lie algebra structure vectors.
\end{definition*}

We refer the reader to~\cite{Jacobson1962} for the basic definitions and properties of Lie algebras.

The properties of the Lie bracket ensure that $\LnF$ is an algebraic subset of $\F^m$.
This is because $\LnF={\bf V}(S)$ where $S$ is the union of the following three subsets of  $\F[X_{ijk}:\ 1\le i,j,k\le n]$
(see, for example,~\cite[pages 4--5]{Jacobson1962} for a proof of this fact):
\begin{gather*}
\{X_{iik}:\ 1\le i,k\le n\},\\
\{X_{ijk}+X_{jik}:\  1\le i,j,k\le n\},\\
\Big\{ \sum_k(X_{ijr}X_{klr}+X_{jlk}X_{kir}+X_{lik}X_{kjr}):\ 1\le i,j,l,r\le n \Big\}.
\end{gather*}

\begin{remark}\label{RemarkNAturalAction}
We have the following natural action of $G=\GlnF$ on ${\mathcal L}_n(\F)$ by `change of basis'.
Let $g=(g_{ij})\in G$ and let $\boldsymbol\lambda=(\lambda_{ijk})\in\LnF$.
Also let $\mathfrak{g}$ be an $n$-dimensional Lie algebra over $\F$ and $\hat b=(b_1,\ldots,b_n)$ an ordered $\F$-basis of $\mathfrak{g}$ such that $\boldsymbol\lambda=(\lambda_{ijk})$ is the structure vector of $\mathfrak{g}$ relative to $\hat b$.
Now let $\hat b'=(b_1',\ldots,b_n')$ be the basis of $\mathfrak{g}$ defined by $b_j'=\sum_{i=1}^ng_{ij}b_i$ for $1\le j\le n$.
Also let $\boldsymbol\lambda'=(\lambda_{ijk}')\in\F^m$ be the structure vector of $\mathfrak{g}$ relative to $\hat b'$
(so we have $[b_i',b_j']=\sum_{k=1}^n\lambda_{ijk}'b_k'$ for $1\le i,j\le n$).
We will write $\boldsymbol\lambda'=\boldsymbol\lambda g$
(clearly, $\boldsymbol\lambda'\in\LnF$).
We call $g$ the transition matrix from basis $\hat b$ to basis $\hat b'$ of $\mathfrak{g}$.

It is well-known and easy to check, that the above process describes a well-defined (right) action of $G$ on $\LnF$.
(See, for example,~\cite[Remark~2.6]{IvanovaPallikaros2017} where some details of such a check are given.)
\end{remark}

Observe that the orbits relative to the action defined in the preceding remark correspond precisely to the isomorphism classes of $n$-dimensional Lie algebras over $\F$.
We denote by $O(\boldsymbol\mu)$ the orbit of the Lie algebra structure vector $\boldsymbol\mu\in\LnF$ under the action of $\GlnF$ described above.

\begin{example*}
It is immediate that the zero vector $\boldsymbol0=(0_{\F},\ldots,0_{\F})$ of $\F^{n^3}$ belongs to ${\mathcal L}_n(\F)$
as it corresponds to the $n$-dimensional Abelian Lie algebra over ${\F}$ (under any choice of basis).
Its orbit consists of precisely one point and hence it is Zariski-closed.
\end{example*}

\begin{remark}\label{RemarkOnMapPhi}
(i) For each $g\in \GlnF$, making use of the action described in Remark~\ref{RemarkNAturalAction}, we define a function $\Phi_g: {\mathcal L}_n(\F)\to {\mathcal L}_n(\F)$:
$\boldsymbol\mu\mapsto \boldsymbol\mu g$, ($\boldsymbol\mu\in{\mathcal L}_n(\F)$).
Then $\Phi_g$ is a regular map and hence continuous in the Zariski topology.
(To see this we fix $g\in \GlnF$. It follows from the change of basis process that for each $\boldsymbol\mu\in{\mathcal L}_n(\F)$ we get
$\Phi_g(\boldsymbol\mu)=(\ev_{\boldsymbol\mu}(f_1),\ldots,\ev_{\boldsymbol\mu}(f_{n^3}))$ where, for $1\le i\le n^3$, $f_i$ is polynomial in $\F[X]$ which only depends on $g$.)

(ii) In view of item (i), one can give an elementary proof of the fact that the closure of an orbit in ${\mathcal L}_n(\F)$ is a union of orbits (see, for example,~\cite[Lemma 3.1]{IvanovaPallikaros2017}).
\end{remark}

\begin{definition*}\label{DefinitionDegeneration}
Let $\mathfrak{g}_1$, $\mathfrak{g}_2$ be $n$-dimensional Lie algebras over $\F$.
We say that $\mathfrak{g}_1$ degenerates to $\mathfrak{g}_2$ (respectively, $\mathfrak{g}_1$ properly degenerates to $\mathfrak{g}_2$)
if there exist structure vectors $\boldsymbol\lambda_1$ of $\mathfrak{g}_1$ and $\boldsymbol\lambda_2$ of $\mathfrak{g}_2$, relative to some bases of $\mathfrak{g}_1$ and $\mathfrak{g}_2$, such that $\boldsymbol\lambda_2\in\overline{O(\boldsymbol\lambda_1)}$
(respectively, $\boldsymbol\lambda_2\in\overline{O(\boldsymbol\lambda_1)}\setminus O(\boldsymbol\lambda_1)$).
\end{definition*}

It is immediate from Remark~\ref{RemarkOnMapPhi}(ii) that if $\boldsymbol\lambda\in\overline{O(\boldsymbol\mu)}$ and $\boldsymbol\nu\in\overline{O(\boldsymbol\lambda)}$, then $\boldsymbol\nu\in\overline{O(\boldsymbol\mu)}$, ($\boldsymbol\lambda,\boldsymbol\mu,\boldsymbol\nu\in\mathcal L_n(\F)$).
In other words, the transitivity property holds in the case of degenerations.

\bigskip
Finally for this subsection we remark that there are no proper degenerations over finite fields as finite subsets of $\F^m$ are closed in the Zariski topology.
%
%

\section{Lie algebra orbit closures via polynomial equations}\label{SectionDegenExamples}

We continue with our assumption that $\F$ is an arbitrary infinite field.

Below, $\mathfrak{h}_3$ will denote the Heisenberg (Lie) algebra, $\mathfrak{g}_2$ will denote the 2-dimensional non-Abelian Lie algebra and $\mathfrak{a}_k$,  for $k\ge1$, the Abelian Lie algebra of dimension $k$.

We will be making use of the action of $G=\GlnF$ on $\LnF$ described in Remark~\ref{RemarkNAturalAction} in order to perform some explicit computations concerning the orbits (and their closure in the Zariski topology) of certain given Lie algebra structure vectors corresponding to $\mathfrak{h}_3$ and $\mathfrak{g}_2\oplus\mathfrak{a}_1$ respectively.
This will allow us to give descriptions of the orbit closures of these structure vectors as algebraic sets (via polynomial equations) and, in addition, obtain descriptions of the particular orbits we investigate here as intersections of a Zariski-closed with a Zariski-open set.

We will also show how these explicit descriptions of the orbits enable us to provide an alternative way of obtaining all degenerations of $\mathfrak{h}_3$ and $\mathfrak{g}_2\oplus\mathfrak{a}_1$ over $\F$.

\subsection{The Heisenberg algebra}\label{SubsectionDegenHeis}

We consider the Heisenberg algebra $\mathfrak{h}_3$.
This (3-dimensional) algebra has an $\F$-basis $\hat e=(e_1,e_2,e_3)$ relative to which the only non-zero products (commutation relations) are  $[e_2,e_3]=e_1=-[e_3,e_2]$.
The structure vector of $\mathfrak{h}_3$ relative to $\hat e$ is $\boldsymbol\eta=(\eta_{ijk})\in\F^{27}$ where $\eta_{231}$ and $\eta_{321}$ (with $\eta_{231}=1$, $\eta_{321}=-1$) are the only nonzero coefficients of~$\boldsymbol\eta$.
First we determine $O(\boldsymbol\eta)$ as a subset of $\F^{27}$.
For this, let $g=(g_{ij})\in \GlF$ and suppose that $M_{ij}$ ($i,j=1,2,3$) is the determinant of the matrix obtained from $g$ by deleting its $i$-th row and $j$-th column.
Assume further that $g$ is the transition matrix from basis $(e_1,e_2,e_3)$ to the basis $(e_1',e_2',e_3')$ of $\mathfrak{h}_3$.
(So $(e_1',e_2',e_3')$ is the basis of $\mathfrak{h}_3$ given by $e_j'=\sum_{i=1}^3g_{ij}e_i$ for $1\le j\le 3$.)
An easy computation then shows that, relative to this new basis, the multiplication in $\mathfrak{h}_3$ is given by
\begin{gather*}
[e_1',e_2']=(\det g)^{-1}M_{13}(M_{11}e_1'-M_{12}e_2'+M_{13}e_3') ,\\
[e_1',e_3']=(\det g)^{-1}M_{12}(M_{11}e_1'-M_{12}e_2'+M_{13}e_3') ,\\
[e_2',e_3']=(\det g)^{-1}M_{11}(M_{11}e_1'-M_{12}e_2'+M_{13}e_3').
\end{gather*}
It follows that there exist $\alpha,\beta,\gamma,\delta\in\F$ such that
\begin{gather*}
[e_1',e_2']=\gamma\delta(\alpha e_1'-\beta e_2'+\gamma e_3'),\\
[e_1',e_3']=\beta\delta(\alpha e_1'-\beta e_2'+\gamma e_3'),\\
[e_2',e_3']=\alpha\delta(\alpha e_1'-\beta e_2'+\gamma e_3').
\end{gather*}
The above relations motivate the following definition.
For $\alpha,\beta,\gamma,\delta\in\F$, let $\boldsymbol\eta'(\alpha,\beta,\gamma,\delta)\in\F^{27}$ be defined by
$\boldsymbol\eta'(\alpha,\beta,\gamma,\delta)=
(0,\,0,\,0,$ $\alpha\gamma\delta,\,-\beta\gamma\delta,\,\gamma^2\delta,$    $\alpha\beta\delta,\,-\beta^2\delta,\,\beta\gamma\delta,$      $-\alpha\gamma\delta,\,\beta\gamma\delta,\,-\gamma^2\delta,$   $0,\,0,\,0,$  $\alpha^2\delta,\,-\alpha\beta\delta,\,\alpha\gamma\delta,$
$ -\alpha\beta\delta,\,\beta^2\delta,\,-\beta\gamma\delta,$  $-\alpha^2\delta,\,\alpha\beta\delta,\,-\alpha\gamma\delta,$   $0,\,0,\,0).
$

We aim to show that the subset $V$ of $\F^{27}$ defined by $V=\{\boldsymbol\eta'(\alpha,\beta,\gamma,\delta)\in\F^{27}:\ \alpha,\beta,\gamma,\delta\in\F\}$ is in fact the (disjoint) union of $O(\boldsymbol\eta)$ and $O(\boldsymbol0)$
(recall that $\boldsymbol0$, the zero vector of $\F^{27}$, is the unique structure vector corresponding to the 3-dimensional Abelian Lie algebra).
It is clear from the above discussion that $O(\boldsymbol\eta)\subseteq V$, hence it suffices to show that any nonzero vector $\boldsymbol v\in V$ belongs to $O(\boldsymbol\eta)$.
For this, it will be convenient to consider the decomposition $V=V_1\cup V_2\cup V_3$ where the subsets $V_1$, $V_2$, $V_3$ of $V$ are defined as follows:
First, for $\mu$, $\nu$, $\lambda$, $\sigma$, $\tau$, $\kappa\in\F$, define the elements $\boldsymbol\eta_1(\mu,\nu,\lambda)$, $\boldsymbol\eta_2(\tau,\sigma)$ and $\boldsymbol\eta_3(\kappa)$ of $\F^{27}$ by
\begin{gather*}
\boldsymbol\eta_1(\mu,\nu,\lambda)=(0,0,0,   \nu\lambda,-\mu\nu\lambda,\nu^2\lambda,  \mu\lambda,-\mu^2\lambda,\mu\nu\lambda,    -\nu\lambda,\mu\nu\lambda,-\nu^2\lambda,   0,0,0,
\lambda,-\mu\lambda,\nu\lambda, \\
\phantom{\boldsymbol\eta_1(\mu,\nu,\lambda)=}{}  -\mu\lambda,\mu^2\lambda,-\mu\nu\lambda,  -\lambda,\mu\lambda,-\nu\lambda,   0,0,0),\\
\boldsymbol\eta_2(\tau,\sigma)=(0,0,0,   0,\sigma\tau,-\sigma\tau^2,    0,\sigma, -\sigma\tau,     0,-\sigma\tau,\sigma\tau^2,   0,0,0,    0,0,0,
0,-\sigma, \sigma\tau, 0,0,0,    0,0,0),\\
\boldsymbol\eta_3(\kappa)=(0,0,0,   0,0,\kappa,    0,0,0,   0,0,-\kappa,    0,0,0,   0,0,0,   0,0,0,   0,0,0,   0,0,0).
\end{gather*}
We then let $V_1=\{\boldsymbol\eta_1(\mu,\nu,\lambda):\ \mu,\nu,\lambda\in\F\}$, $V_2=\{\boldsymbol\eta_2(\tau,\sigma):\ \tau,\sigma\in\F\}$ and $V_3=\{\boldsymbol\eta_3(\kappa): \kappa\in\F\}$.

In order to establish that $V$ is indeed the union of the three sets above, it suffices to verify that
$V_1=\{\boldsymbol\eta'(\alpha,\beta,\gamma,\delta)\in V:\ \alpha\ne0\}$, $V_2=\{\boldsymbol\eta'(\alpha,\beta,\gamma,\delta)\in V:\ \alpha=0$ and $\beta\ne0\}$ and $V_3=\{\boldsymbol\eta'(\alpha,\beta,\gamma,\delta)\in V:\ \alpha=0$ and $\beta=0\}$.
That the above equalities of sets in fact hold is immediate from the relations $\boldsymbol\eta'(1,\mu,\nu,\lambda)=\boldsymbol\eta_1(\mu,\nu,\lambda)$, $\boldsymbol\eta_1(\beta\alpha^{-1},\gamma\alpha^{-1},\delta\alpha^2)=\boldsymbol\eta'(\alpha,\beta,\gamma,\delta)$ (for $\alpha\ne0$), $\boldsymbol\eta'(0,1,\tau,-\sigma)=\boldsymbol\eta_2(\tau,\sigma)$, $\boldsymbol\eta_2(\gamma\beta^{-1},-\delta\beta^2)=\boldsymbol\eta'(0,\beta,\gamma,\delta)$ (for $\beta\ne0$) and $\boldsymbol\eta'(0,0,1,\kappa)=\boldsymbol\eta_3(\kappa)$, $\boldsymbol\eta_3(\delta\gamma^2)=\boldsymbol\eta'(0,0,\gamma,\delta)$.

Since $V=V_1\cup V_2\cup V_3$, we can see that any nonzero element of $V$ has one of the following forms: $\boldsymbol\eta_1(\mu,\nu,\lambda)$ (with $\lambda\ne0$), $\boldsymbol\eta_2(\tau,\sigma)$ (with $\sigma\ne0$) or $\boldsymbol\eta_3(\kappa)$ (with $\kappa\ne0$).

Moreover, for $\lambda\ne0$ we have $\boldsymbol\eta g_1(\mu,\nu,\lambda)=\boldsymbol\eta_1(\mu,\nu,\lambda)$, for $\sigma\ne0$ we have $\boldsymbol\eta g_2(\tau,\sigma)=\boldsymbol\eta_2(\tau,\sigma)$ and finally for $\kappa\ne0$ we have $\boldsymbol\eta g_3(\kappa)=\boldsymbol\eta_3(\kappa)$ where, for $\lambda\ne0$, $\sigma\ne0$, $\kappa\ne0$ respectively, the matrices
\[
g_1(\mu,\nu,\lambda)=
\left[
\begin{array}{lll}
\lambda^{-1} & 0 & 0\\
\mu & 1 & 0\\
-\nu & 0 & 1
\end{array}
\right],
\quad
g_2(\tau,\sigma)=
\left[
\begin{array}{lll}
0 & \sigma^{-1} & 0\\
1 & 0 & 0\\
0 & \tau & 1
\end{array}
\right],
\quad
g_3(\kappa)=
\left[
\begin{array}{lll}
0 & 0 & \kappa^{-1}\\
1 & 0 & 0\\
0 & 1 & 0
\end{array}
\right]
\]
all belong to $G=\GlF$.
This establishes that $V\setminus\{\boldsymbol0\}=O(\boldsymbol\eta)$.

Our next aim is to show that $V$ is an irreducible algebraic set.
For this, let $S=S_1\cup S_2\cup S_3$ where $S_1$, $S_2$, $S_3$ are the following subsets of $\F[X_{ijk}:\ 1\le i,j,k\le 3]$:
\begin{gather*}
S_1=\{X_{iik}:\ 1\le i,k\le 3\},\qquad S_2=\{X_{ijk}+X_{jik}:\ 1\le i,j,k\le 3\},\\
S_3=\{X_{121}-X_{233},\quad X_{131}+X_{232},\quad X_{122}+X_{133},\quad X_{122}^2+X_{123}X_{132},\\
X_{121}^2-X_{123}X_{231},\quad  X_{131}^2+X_{132}X_{231},\quad  X_{121}X_{131}+X_{122}X_{231}\}.
\end{gather*}
Observe that $S\subseteq {\bf I}(V)$.
We claim that $V={\bf V}(S)$.
It is clear that $V\subseteq {\bf V}(S)$.
To establish the reverse inclusion ${\bf V}(S)\subseteq V$, let $\boldsymbol\gamma=(\gamma_{ijk})\in\F^{27}$ be a common zero of the elements of~$S$.
Since $\boldsymbol\gamma$ is a common zero of the elements of~$S_1\cup S_2$, we see that the shape of $\boldsymbol\gamma$ is determined once we determine the shape of the auxiliary vector $\hat{\boldsymbol\gamma} =(\gamma_{121},\gamma_{122},\gamma_{123},\gamma_{131},\gamma_{132},\gamma_{133},\gamma_{231},\gamma_{232},\gamma_{233})\in\F^9$.
Invoking now the fact that $\boldsymbol\gamma$ is a common zero of the polynomials of degree $1$ in $S_3$ we see that in fact $\hat{\boldsymbol\gamma}$ has shape $(\gamma_{121},\gamma_{122},\gamma_{123},\gamma_{131},\gamma_{132},-\gamma_{122},\gamma_{231},-\gamma_{131},\gamma_{121})$.
We will consider the cases $\gamma_{231}\ne0$ and $\gamma_{231}=0$ separately.
If $\lambda=\gamma_{231}\ne0$ we can set $\mu=\gamma_{131}\lambda^{-1}$ and $\nu=\gamma_{121}\lambda^{-1}$ from which we can deduce that $\gamma_{123}=\nu^2\lambda$ (since $\gamma_{121}^2-\gamma_{123}\gamma_{231}=0$), $\gamma_{132}=-\mu^2\lambda$ (since $\gamma_{131}^2+\gamma_{132}\gamma_{231}=0$) and $\gamma_{122}=-\mu\nu\lambda$ (since $\gamma_{121}\gamma_{131}+\gamma_{122}\gamma_{231}=0$).
Hence $\boldsymbol\gamma\in V_1$ whenever $\gamma_{231}\ne0$.

For the case $\gamma_{231}=0$, by similar argument, one can show that if $\gamma_{132}\ne0$, then $\boldsymbol\gamma\in V_2$ and if $\gamma_{132}=0$ then $\boldsymbol\gamma\in V_3$.
We conclude that $V={\bf V}(S)$ and hence $V$ is an algebraic set.

Next, we consider the map $\Phi:\F^4\to\F^{27}$: $(\alpha,\beta,\gamma,\delta)\mapsto\boldsymbol\eta'(\alpha,\beta,\gamma,\delta)$.
Clearly $\Phi$ is a regular map having $V$ as its image.
Thus, $\overline{\Phi(\F^4)}=\overline{V}=V$.
Invoking Remarks~\ref{Remark1.1.13Geck} and~\ref{Remark1.3.2Geck}, we see that $V$ is irreducible.
It follows from this that $O(\boldsymbol\eta)$ is not closed in the Zariski topology.
(Note that if $O(\boldsymbol\eta)$ were Zariski-closed this would imply that $V=O(\boldsymbol\eta)\cup\{\boldsymbol0\}$ is reducible, being the union of two nonempty closed sets.)
Hence, $O(\boldsymbol\eta)$ is properly contained in $\overline{O(\boldsymbol\eta)}$.
Also $\overline{O(\boldsymbol\eta)}\subseteq V$ since ${O(\boldsymbol\eta)}\subseteq V$ and $V$ is an algebraic set.
We conclude that $\overline{O(\boldsymbol\eta)}=V=O(\boldsymbol\eta)\cup\{\boldsymbol0\}$.
In other words, over an arbitrary infinite field, the only proper degeneration of $\mathfrak{h}_3$ is to the Abelian Lie  algebra $\mathfrak a_3$.
We remark here that this is a well-known fact and has been proved using different methods over various fields, see for example~\cite{Conatser1972,Gorbatsevich1991,IvanovaPallikaros2017,Sharp1960}.
In the discussion above we presented an alternative way of obtaining it, based on the definition of an irreducible algebraic set.

\subsection{The algebra $\mathfrak{g}_2\oplus\mathfrak{a}_1$}\label{SubsectionDegenA21}

In this subsection we perform a similar investigation for the algebra $\mathfrak{g}=\mathfrak{g}_2\oplus\mathfrak{a}_1$.
Note that this algebra has an $\F$-basis $\hat b=(b_1,b_2,b_3)$ relative to which the only non-zero commutation relations are given by $[b_1,b_2]=b_1=-[b_2,b_1]$.
Let $\boldsymbol\rho=(\rho_{ijk})\in\F^{27}$ be the structure vector of $\mathfrak{g}$ relative to the basis $\hat b$.
Suppose now that $g\in G=\GlF$ is the transition matrix from $\hat b$ to the basis $\hat b'=(b_1',b_2',b_3')$ of $\mathfrak{g}$.
It is then easy to show that
\begin{gather*}
[b_1',b_2']=(\det g)^{-1}{M_{33}}(M_{11}b_1'-M_{12}b_2'+M_{13}b_3'),\\
[b_1',b_3']=(\det g)^{-1}{M_{32}}(M_{11}b_1'-M_{12}b_2'+M_{13}b_3'),\\
[b_2',b_3']=(\det g)^{-1}{M_{31}}(M_{11}b_1'-M_{12}b_2'+M_{13}b_3')
\end{gather*}
where, as before, $M_{ij}$ denotes the determinant of the matrix obtained from $g$ by deleting its  $i$-th row and $j$-th column
(in particular, the $M_{ij}$ are elements of our field $\F$).
It follows that there exist $\chi_1,\psi_1,\omega_1,\chi_2,\psi_2,\omega_2,\delta\in\mathbb \F$ such that
\begin{gather*}
[b_1',b_2']=\delta{\chi_2}(\chi_1b_1'-\psi_1b_2'+\omega_1b_3'),\\
[b_1',b_3']=\delta{\psi_2}(\chi_1b_1'-\psi_1b_2'+\omega_1b_3'),\\
[b_2',b_3']=\delta{\omega_2}(\chi_1b_1'-\psi_1b_2'+\omega_1b_3').
\end{gather*}

This prompts us to define $\boldsymbol\rho'(\chi_1,\psi_1,\omega_1,\chi_2,\psi_2,\omega_2,\delta)\in\F^{27}$ by $\boldsymbol\rho'(\chi_1,\psi_1,\omega_1,\chi_2,\psi_2,\omega_2,\delta)=(0,\,0,\,0,$    $\chi_1\chi_2\delta,$ $-\psi_1\chi_2\delta,$ $\omega_1\chi_2\delta,$     $\chi_1\psi_2\delta,$ $-\psi_1\psi_2\delta,\,\omega_1\psi_2\delta,$   $-\chi_1\chi_2\delta,\,\psi_1\chi_2\delta,\,-\omega_1\chi_2\delta,$   $0,\,0,\,0,$ $\chi_1\omega_2\delta,$ $-\psi_1\omega_2\delta,$ $\omega_1\omega_2\delta,$   $-\chi_1\psi_2\delta,\,\psi_1\psi_2\delta,$ $-\omega_1\psi_2\delta,$   $-\chi_1\omega_2\delta,$ $\psi_1\omega_2\delta,$ $-\omega_1\omega_2\delta,$   $0,\,0,\,0)$,
and the subset $U$ of $\F^{27}$ by $U=\{\boldsymbol\rho'(\chi_1,\psi_1,\omega_1,\chi_2,\psi_2,\omega_2,\delta):\ \chi_1,\psi_1,\omega_1,\chi_2,\psi_2,\omega_2,\delta\in\mathbb \F\}$.

It is then clear that $O(\boldsymbol\rho)\subseteq U$.
We want to show that $U$ is an algebraic set containing $V=O(\boldsymbol\eta)\cup\{\boldsymbol0\}$ (we keep the notation for $V$, $\boldsymbol\eta$, $\boldsymbol\eta'$ and also for $S$, $S_1$, $S_2$, $S_3$ introduced in the previous subsection).
The inclusion $V\subseteq U$ is immediate from the fact that $\boldsymbol\eta'(\alpha,\beta,\gamma,\delta)=\boldsymbol\rho'(\alpha,\beta,\gamma, \gamma, \beta, \alpha, \delta)$.

Next, we define the subset $T$ of ${\bf I}(U)$ by $T=S_1\cup S_2\cup T_3$ where
\begin{gather*}
T_3=\{X_{121}X_{132}-X_{122}X_{131},\quad  X_{121}X_{232}-X_{122}X_{231},\quad  X_{131}X_{232}-X_{132}X_{231},\\
X_{121}X_{133}-X_{123}X_{131},\quad  X_{121}X_{233}-X_{123}X_{231},\quad  X_{232}X_{123}-X_{122}X_{233},\\
X_{122}X_{133}-X_{123}X_{132},\quad  X_{132}X_{233}-X_{133}X_{232},\quad  X_{233}X_{131}-X_{133}X_{231}\}
\end{gather*}
(recall the definition of $S_1$ and $S_2$ in Subsection~\ref{SubsectionDegenHeis}).

Now let $S'=T\cup \{X_{121}-X_{233},\  X_{131}+X_{232},\  X_{122}+X_{133}\}\subseteq T\cup S_3$.
It is easy to check that $V\subseteq{\bf V}(S')$.
We also have ${\bf V}(S')={\bf V}(T\cup S_3)$.
To see this last equality of sets, note first that ${\bf V}(T\cup S_3)\subseteq{\bf V}(S')$ since $S'\subseteq T\cup S_3$.
On the other hand, any $\boldsymbol\nu\in{\bf V}(S')$ is a common zero of every polynomial in $T\cup S_3$.
Hence, we also have ${\bf V}(S')\subseteq{\bf V}(T\cup S_3)$.
Since $V\subseteq{\bf V}(S')$, we get $V\subseteq{\bf V}(T\cup S_3)$.
But $T\cup S_3\supseteq S$, so ${\bf V}(T\cup S_3)\subseteq{\bf V}(S)=V$.
We conclude that $V\,(={\bf V}(S))={\bf V}(T\cup S_3)={\bf V}(S')$.

We aim to show that $U={\bf V}(T)$.
This would imply that $U$ is an algebraic set (and also provide an alternative way of seeing that $V\subseteq U$ in view of the observation above).

Clearly, $U\subseteq{\bf V}(T)$.
In order to establish the reverse inclusion, it will be convenient to decompose $U$ as a union of three subsets which contain among them all elements of ${\bf V}(T)$.
With $\alpha$, $\beta$, $\gamma$, $\mu$, $\nu$, $\phi$, $\rho$, $\sigma$, $\tau$, $\zeta$, $\theta$, $\xi$, $\kappa\in\F$, define the elements $\boldsymbol\rho_1(\alpha,\beta,\gamma,\mu,\nu,\phi)$, $\boldsymbol\rho_2(\sigma,\tau,\rho,\zeta)$ and $\boldsymbol\rho_3(\theta,\xi,\kappa)\in\F^{27}$ by
\begin{gather*}
\boldsymbol\rho_1(\alpha,\beta,\gamma,\mu,\nu,\phi)=(0,0,0,   \mu\alpha,-\mu\beta,\mu\gamma,  \nu\alpha,-\nu\beta,\nu\gamma,    -\mu\alpha,\mu\beta,-\mu\gamma,   0,0,0,
\phi\alpha,-\phi\beta,\phi\gamma, \\
\phantom{S_1(\alpha,\beta,\gamma,\mu,\nu,\phi)=(}{}  -\nu\alpha,\nu\beta,-\nu\gamma,  -\phi\alpha,\phi\beta,-\phi\gamma,   0,0,0),\\
\boldsymbol\rho_2(\sigma,\tau,\rho,\zeta)=(0,0,0,   0,\sigma,-\sigma \zeta,    0,\tau,-\tau \zeta,      0,-\sigma,\sigma \zeta,   0,0,0,    0,\rho,-\rho \zeta, 0,-\tau,\tau \zeta,     0,-\rho,\rho \zeta,  \\
\phantom{S_1(\sigma,\tau,\rho,\zeta)=(}{}  0,0,0),\\
\boldsymbol\rho_3(\theta,\xi,\kappa)=(0,0,0,   0,0,\theta,    0,0,\xi,   0,0,-\theta,    0,0,0,   0,0,\kappa,   0,0,-\xi,   0,0,-\kappa,   0,0,0).
\end{gather*}
Also define the subsets $U_1$, $U_2$ and $U_3$ of $\F^{27}$ by $U_1=\{\boldsymbol\rho_1(\alpha,\beta,\gamma,\mu,\nu,\phi):\ \alpha,\beta,\gamma,\mu,\nu,\phi\in\F$ and $\alpha\ne0\}$,
$U_2=\{\boldsymbol\rho_2(\sigma,\tau,\rho,\zeta):$ $\sigma,\tau,\rho,\zeta\in\F\}$ and
$U_3=\{\boldsymbol\rho_3(\theta,\xi,\kappa):$ $\theta,\xi,\kappa\in\F\}$.

It is then immediate from the relations
$\boldsymbol\rho_1(\alpha,\beta,\gamma,\mu,\nu,\phi) = \boldsymbol\rho'(\chi_1=\alpha, \psi_1=\beta, \omega_1=\gamma, \chi_2=\mu, \psi_2=\nu, \omega_2=\phi, \delta=1)$,
$\boldsymbol\rho_2(\sigma,\tau,\rho,\zeta)=\boldsymbol\rho'(\chi_1=0, \psi_1=-1, \omega_1=-\zeta, \chi_2=\sigma, \psi_2=\tau, \omega_2=\rho, \delta=1)$
and $\boldsymbol\rho_3(\theta,\xi,\kappa)=\boldsymbol\rho'(\chi_1=0, \psi_1=0, \omega_1=1, \chi_2=\theta, \psi_2=\xi, \omega_2=\kappa, \delta=1)$
that $U_i\subseteq U$ for $i=1,2,3$.

We now show that ${\bf V}(T)\subseteq U_1\cup U_2\cup U_3$.
Let $\boldsymbol\gamma=(\gamma_{ijk})\in\F^{27}$ be a common zero of all polynomials in $T$.
As $T\supseteq S_1\cup S_2$, similarly to the Heisenberg algebra case, we will work with the auxiliary vector $\hat{\boldsymbol\gamma}=(\gamma_{121}, \gamma_{122}, \gamma_{123}, \gamma_{131}, \gamma_{132}, \gamma_{133}, \gamma_{231}, \gamma_{232}, \gamma_{233})\in\F^9$.
Again, we will need to consider different subcases.
We begin by considering the case  $\gamma_{121}\ne0$.
Since $\boldsymbol\gamma\in{\bf V}(T)$, we get $\hat{\boldsymbol\gamma}=(\gamma_{121}, \gamma_{122}, \gamma_{123}, \gamma_{131}, \gamma_{122}\gamma_{131}\gamma_{121}^{-1}, \gamma_{123}\gamma_{131}\gamma_{121}^{-1}, \gamma_{231}, \gamma_{122}\gamma_{231}\gamma_{121}^{-1}, \gamma_{123}\gamma_{231}\gamma_{121}^{-1})$.
For example, to see that $\gamma_{132}=\gamma_{122}\gamma_{131}\gamma_{121}^{-1}$, note that $\boldsymbol\gamma$ is a zero of the polynomial $X_{121}X_{132}-X_{122}X_{131}$ which belongs to $T$.
On setting $\mu=1$, $\nu=\gamma_{131}\gamma_{121}^{-1}$, $\phi=\gamma_{231}\gamma_{121}^{-1}$, $\alpha=\gamma_{121}\,(\ne0)$, $\beta=-\gamma_{122}$, $\gamma=\gamma_{123}$, we see that $\boldsymbol\gamma=\boldsymbol\rho_1(\alpha,\beta,\gamma,\mu,\nu,\phi)$ where $\alpha\ne0$, so $\boldsymbol\gamma\in U_1$.
Next we consider the case $\gamma_{121}=0$.
We split this case into the subcases $\gamma_{122}\ne0$ (where, by similar argument as above, we can show that $\boldsymbol\gamma\in U_2$) and $\gamma_{122}=0$.
It remains to consider the case when when both $\gamma_{121}$ and $\gamma_{122}$ are equal to zero and the next step is to split this case into subcases according to whether $\gamma_{123}\ne0$ (we can show then that $\boldsymbol\gamma\in U_3$) or $\gamma_{123}=0$.
Continuing in a similar fashion, we finally deduce that ${\bf V}(T)$ is indeed a subset of $U_1\cup U_2\cup U_3$.
Summing up the above discussion, we see that $U\subseteq {\bf V}(T)\subseteq U_1\cup U_2\cup U_3\subseteq U$.
This forces $U=U_1\cup U_2\cup U_3={\bf V}(T)$.
Recall now that $V=O(\boldsymbol\eta)\cup\{\boldsymbol0\}\subseteq U$.
In order to show that $U=O(\boldsymbol\rho)\cup O(\boldsymbol\eta)\cup\{\boldsymbol0\}$, we find, for each $\boldsymbol\delta\in U\setminus V$, an invertible matrix $g({\boldsymbol\delta})\in G$ such that $\boldsymbol\delta=\boldsymbol\rho\, g({\boldsymbol\delta})$.

In the table below we summarize the results of this computation, listing also the corresponding matrices $g=g({\boldsymbol\delta})$.
We first split into subcases according to whether $\boldsymbol\delta\in U\setminus V$ is of the form $\boldsymbol\rho_1$ (with $\alpha\ne0$), $\boldsymbol\rho_2$ or $\boldsymbol\rho_3$ and as it turns out, depending on the values of the elements of $\F$ involved, we need to split into further subcases.

It is now useful to recall that $V={\bf V}(S')$ where $S'=T\cup \{X_{121}-X_{233},\  X_{131}+X_{232},\  X_{122}+X_{133}\}\subseteq T\cup S_3$.
Let $\boldsymbol\rho'=(\rho'_{ijk})\in U$.
It follows that $\boldsymbol\rho'\in V$ if, and only if all three conditions $\rho'_{121}-\rho'_{233}=0$, $\rho'_{131}+\rho'_{232}=0$ and $\rho'_{122}+\rho'_{133}=0$ are satisfied.
In particular, in the case $\boldsymbol\rho'=\boldsymbol\rho_1(\alpha,\beta,\gamma,\mu,\nu,\phi)$, we have $\boldsymbol\rho'\in V$ if, and only if, all of the conditions $\mu\alpha-\phi\gamma=0$, $\nu\alpha-\phi\beta=0$ and $-\mu\beta+\nu\gamma=0$ are satisfied.
For $\boldsymbol\rho_1$ to be an element of $U_1$ we have the restriction $\alpha\ne0$, so in this case, the third of the last three conditions follows from the other two
(this is because the conditions $\mu\alpha-\phi\gamma=0$ and $\nu\alpha-\phi\beta=0$ are equivalent to the conditions $\mu=\phi\gamma\alpha^{-1}$ and $\nu=\phi\beta\alpha^{-1}$ if $\alpha\ne0$).
For simplicity, in the table below we will be writing $A_1=\mu\alpha-\phi\gamma$, $A_2=\nu\alpha-\phi\beta$.
Similar observations can be made in the cases $\boldsymbol\rho'$ has form $\boldsymbol\rho_2$ or $\boldsymbol\rho_3$ (as it can also be seen from the table).
Moreover, in the table below, vector $\boldsymbol\rho_1=\boldsymbol\rho_1(\alpha,\beta,\gamma,\mu,\nu,\phi)$ will always be considered under the restriction $\alpha\ne0$, compare with the definition of set $U_1$.

{\footnotesize
\begin{longtable}{|l|l|l|l|}
\hline
$\boldsymbol\rho_i$ & conditions & transition matrix $g$ & $\det g$\\
\hline
$\boldsymbol\rho_1$& $\begin{array}{l}A_1\ne0,\ A_2\ne0\end{array}$ &
$\left[
\begin{array}{lll}
\nu & \phi & 0\\
\mu\beta-\nu\gamma & A_1 & A_2\\
-\gamma & 0 & \alpha
\end{array}
\right]$ & $A_1A_2$ \\
\hline
$\boldsymbol\rho_1$ & $\begin{array}{l}A_1=0,\  A_2\ne0,\\ \phi\gamma\ne0\end{array}$ &
$
\left[
\begin{array}{lll}
\beta & \alpha & {\alpha}{\phi^{-1}\gamma^{-1}} A_2\\
-{\gamma}\alpha^{-1} A_2 & 0 & A_2\\
{\phi\beta}\alpha^{-1} & \phi & 0
\end{array}
\right]
$ & $-A_2^2$
\\
\hline
$\boldsymbol\rho_1$ &$\begin{array}{l}A_1=0,\  A_2\ne0,\\  \phi=0\\ (\Rightarrow \mu=0,\ \nu\ne0)\end{array}$ & $\left[
\begin{array}{lll}
1 & 0 & 0\\
-\gamma\nu & 0 & \alpha\nu\\
-\gamma+\beta &\alpha  & \alpha
\end{array}
\right]$ & $-\alpha^2\nu$
\\
\hline
$\boldsymbol\rho_1$ &$\begin{array}{l}A_1=0,\  A_2\ne0,\\  \phi\ne0,\ \gamma=0\\ (\Rightarrow \mu=0)\end{array}$& $\left[
\begin{array}{lll}
\nu & \phi & 0\\
0 & 0 & A_2\\
\beta &\alpha  & 0
\end{array}
\right]$ & $-A_2^2$
\\
\hline
$\boldsymbol\rho_1$ &$\begin{array}{l}A_1\ne0,\ A_2=0,\\  \beta\ne0,\  \gamma\ne0\end{array}$ & $\left[
\begin{array}{lll}
0 & \alpha^2\mu\gamma & \alpha\gamma\phi\beta\\
\alpha^{-1}{\beta}A_1 & A_1 & 0\\
{-}\alpha^{-1}A_1 & 0 &\gamma^{-1}{A_1}
\end{array}
\right]$ & $-\beta A_1^3$
\\
\hline
$\boldsymbol\rho_1$ & $\begin{array}{l}A_1\ne0,\ A_2=0,\\  \gamma=0\ (\Rightarrow\mu\ne0) \end{array}$ & $\left[
\begin{array}{lll}
-\mu & 0 & \phi\\
\beta\mu & \mu\alpha & 0\\
\beta & \alpha  & 1
\end{array}
\right]$ & $-\mu^2\alpha$
\\
\hline
$\boldsymbol\rho_1$ &$\begin{array}{l}A_1\ne0,\ A_2=0,\\  \beta=0\ (\Rightarrow\nu=0)\end{array}$ &
$\left[
\begin{array}{lll}
\mu & 0 & -\phi\\
0 & A_1 & 0\\
-\gamma & 0 & \alpha
\end{array}
\right]$
& $A_1^2$
\\
\hline
$\boldsymbol\rho_1$ & $\begin{array}{l}A_1=0,\ A_2=0 \end{array}$ & $\boldsymbol\rho_1\in\overline{O(\boldsymbol\eta)}$ & ---
\\
\hline
$\boldsymbol\rho_2$ & $\rho\ne0$, $\tau \zeta-\sigma\ne0$ & $\left[
\begin{array}{lll}
0 & -\sigma & -\tau\\
\tau \zeta-\sigma & \rho \zeta & \rho\\
1 & 0 & 0
\end{array}
\right]$ & $\rho(\tau \zeta-\sigma)$
\\
\hline
$\boldsymbol\rho_2$ & $\rho\ne0$, $\tau \zeta-\sigma=0$ & $\left[
\begin{array}{lll}
\tau & \rho & 0\\
0 & \rho \zeta & \rho\\
1 & 0 & 0
\end{array}
\right]$ & $\rho^2$
\\
\hline
$\boldsymbol\rho_2$ & $\rho=0$, $\tau \zeta-\sigma\ne0$ & $\left[
\begin{array}{lll}
0 & \sigma & \tau\\
\tau \zeta-\sigma & 0 & 0\\
0 & \zeta & 1
\end{array}
\right]$ & $(\tau \zeta-\sigma)^2$
\\
\hline
$\boldsymbol\rho_2$ & $\rho=0,\ \tau \zeta-\sigma=0$ & $\boldsymbol\rho_2\in\overline{O(\boldsymbol\eta)}$ & ---
\\
\hline
$\boldsymbol\rho_3$ & $\kappa\ne0$, $\xi\ne0$ &$\left[
\begin{array}{lll}
1 & \xi^{-1}{(\kappa+\theta)} & 1\\
-\xi & -\kappa & 0\\
1 & 0 & 0
\end{array}
\right]$ & $\kappa$
\\
\hline
$\boldsymbol\rho_3$ & $\kappa\ne0$, $\xi=0$ &$\left[
\begin{array}{lll}
-\theta \kappa^{-1} & 0 & 1\\
0 & -\kappa & 0\\
1 & 0 & 0
\end{array}
\right]
$ & $\kappa$
\\
\hline
$\boldsymbol\rho_3$ & $\kappa=0$, $\xi\ne0$ & $\left[
\begin{array}{lll}
0 & \theta \xi^{-1} & 1\\
-\xi & 0 & 0\\
0 & 1 & 0
\end{array}
\right]$ & $-\xi$
\\
\hline
$\boldsymbol\rho_3$ & $\kappa=0$, $\xi=0$ & $\boldsymbol\rho_3\in\overline{O(\boldsymbol\eta)}$ & ---
\\
\hline

\end{longtable}
}

\bigskip
The computation above establishes that $U=O(\boldsymbol\rho)\cup O(\boldsymbol\eta)\cup\{\boldsymbol0\}$.
Now recall that $U\,(={\bf V}(T))$ is Zariski-closed.
In fact, by similar argument as in the case of the set $V$, we can show that $U$ is irreducible, considering now the regular map $\Phi:\F^7\to U= \bar U\subseteq F^{27}$: $(\chi_1,\psi_1,\omega_1,\chi_2,\psi_2,\omega_2,\delta)\mapsto (0,\,0,\,0,$    $\chi_1\chi_2\delta,$ $-\psi_1\chi_2\delta,$ $\omega_1\chi_2\delta,$     $\chi_1\psi_2\delta,$ $-\psi_1\psi_2\delta,\,\omega_1\psi_2\delta,$   $-\chi_1\chi_2\delta,\,\psi_1\chi_2\delta,\,-\omega_1\chi_2\delta,$   $0,\,0,\,0,$ $\chi_1\omega_2\delta,$ $-\psi_1\omega_2\delta,$ $\omega_1\omega_2\delta,$   $-\chi_1\psi_2\delta,\,\psi_1\psi_2\delta,$ $-\omega_1\psi_2\delta,$   $-\chi_1\omega_2\delta,$ $\psi_1\omega_2\delta,$ $-\omega_1\omega_2\delta,$   $0,\,0,\,0)=\boldsymbol\rho'(\chi_1,\psi_1,\omega_1,\chi_2,\psi_2,\omega_2,\delta)$.

Since $U$ is Zariski-closed and ${O(\boldsymbol\rho)}\subseteq U$, we get $\overline{O(\boldsymbol\rho)}\subseteq U$.
Invoking the fact that $U=O(\boldsymbol\rho)\cup O(\boldsymbol\eta)\cup\{\boldsymbol0\}$ we can deduce that $\mathfrak{h}_3$ and $\mathfrak{a}_3$ are the only possible Lie algebras which $\mathfrak{g}_2\oplus\mathfrak{a}_1$ can properly degenerate to.
In order to establish that $\mathfrak{g}_2\oplus\mathfrak{a}_1$ in fact degenerates to both $\mathfrak{h}_3$ and $\mathfrak{a}_3$ it suffices to show that $\overline{O(\boldsymbol\rho)}=U$.
Since $U$ is irreducible and $\overline{O(\boldsymbol\eta)}=O(\boldsymbol\eta)\cup\{\boldsymbol0\}$ we get that $O(\boldsymbol\rho)$ is not Zariski-closed.
It follows that $O(\boldsymbol\rho)$ is properly contained in $\overline{O(\boldsymbol\rho)}$.
If $\boldsymbol\eta\not\in\overline{O(\boldsymbol\rho)}$, then $O(\boldsymbol\eta)\cap\overline{O(\boldsymbol\rho)}=\varnothing$ since $\overline{O(\boldsymbol\rho)}$ is a union of orbits
(see Remark~\ref{RemarkOnMapPhi}(ii)).
It would then follow that $\overline{O(\boldsymbol\rho)}=O(\boldsymbol\rho)\cup\{\boldsymbol0\}$, contradicting the fact that $U$ is irreducible.
We conclude that $\boldsymbol\eta\in\overline{O(\boldsymbol\rho)}$.
It follows that $O(\boldsymbol\eta)\subseteq\overline{O(\boldsymbol\rho)}$ and hence $\overline{O(\boldsymbol\eta)}\subseteq\overline{O(\boldsymbol\rho)}$.
Since $\boldsymbol0\in\overline{O(\boldsymbol\eta)}$, we get that $\boldsymbol0\in\overline{O(\boldsymbol\rho)}$.
Summing up, we have shown $\overline{O(\boldsymbol\rho)}\subseteq U=O(\boldsymbol\rho)\cup O(\boldsymbol\eta)\cup\{\boldsymbol0\}\subseteq\overline{O(\boldsymbol\rho)}$.
Hence, $U=\overline{O(\boldsymbol\rho)}$ as required.

We remark here that it is well-known that, over an infinite field, any Lie algebra degenerates to the abelian Lie algebra of the same dimension.
Also note that already in~\cite{Conatser1972} it is shown that $\mathfrak g$ degenerates to $\mathfrak h_3$ in the case the ground field is $\mathbb R$.
In view of~\cite[Lemma~3.9]{IvanovaPallikaros2017} the technique used in~\cite{Conatser1972} can be extended to obtain a degeneration from $\mathfrak g$ to $\mathfrak h_3$ now over an arbitrary infinite field.
In the discussion above we provided an alternative way of obtaining this particular degeneration using the notion of an irreducible algebraic set.

We close this subsection with some general comments regarding our sets above.
First, we can observe that $O(\boldsymbol\rho)=U\setminus\overline{O(\boldsymbol\eta)}=\overline{O(\boldsymbol\rho)}\setminus\overline{O(\boldsymbol\eta)}$ so $O(\boldsymbol\rho)$ is open in its closure (compare~\cite[Proposition~2.5.2]{Geck2003} for the case of an algebraically closed field).
Now let $W$ be the union of the three principal open sets $\{\boldsymbol\alpha\in\F^{n^3}:$ $f_i(\boldsymbol\alpha)\ne0\}$ for $i=1,2,3$ where $f_1=X_{121}-X_{233}$, $f_2=X_{131}+X_{232}$ and $f_3=X_{122}+X_{133}$.
Since $\overline{O(\boldsymbol\rho)}={\bf V}(T)$ and $\overline{O(\boldsymbol\eta)}={\bf V}(S')$  where $S'=T\cup \{f_1,\  f_2,\  f_3\}$, we see that $O(\boldsymbol\rho)={\bf V}(T)\cap W$.
This in fact verifies that ${O(\boldsymbol\rho)}$ consists of precisely those points in $U\,(=\overline{O(\boldsymbol\rho)})$ which do not correspond to unimodular Lie algebras (compare, for example, with~\cite[Remark~4.12]{IvanovaPallikaros2017}).

\end{document}